\documentclass[aps,prb,twocolumn,groupedaddress]{revtex4}

\usepackage{gensymb}
\usepackage{multirow}
\usepackage[dvipdf]{graphicx}
\usepackage{rotating}
\usepackage{natbib}

\begin{document}

\title{Electronic structure at nanocontacts of surface passivated CdSe nanorods with gold clusters}

\author{Deepashri Saraf}
\author{Anjali Kshirsagar}
\email{anjali@cms.unipune.ac.in}
\affiliation{Department of Physics and Center for Modeling and Simulation, University of Pune, Ganeshkhind, Pune 411007, India}
   

\begin{abstract} 
We report the effects of variation in length on the electronic
structure of CdSe nanorods derived from atomic clusters and passivated by
fictitious hydrogen atoms. These nanorods are augmented by attaching gold
clusters at both the ends to form a nanodumbbell. The goal is to assess the
changes at nanolevel after formation of contacts with gold clusters serving as
electrodes and compare the results with experimental observations \cite{SMBM}.
Calculations involving nanorods of length 4.6~\AA\ to 116.6~\AA\ are performed
using density functional theory implemented within plane-wave basis set. The
binding energy per atom saturates for nanorod of length 116.6~\AA.  It is
interesting to note that upon attaching gold clusters, the nanorods shorter
than 27~\AA\ develop metallicity by means of metal induced gap states (MIGS).
Longer nanorods exhibit a nanoscale Schottky barrier emerging at the center.
For these nanorods, interfacial region closest to the gold electrodes shows a
finite density of states in the gap due to MIGS, which gradually decreases
towards the center of the nanorod opening up a finite gap.  Bader charge
analysis indicates localized charge transfer from metal to semiconductor.
\end{abstract}

\maketitle

\section{Introduction}
Group II-VI semiconductors possess an ionic character which leads to larger
band gaps (BGs) and highly coordinated structures in bulk. This property
differentiates them from other compound semiconductors in terms of size and
structure dependent aspects at nanoscale. CdSe being the most popular amongst
them, due to the reproducibility of its optical absorption and emission
properties, has potential applications such as biosensors \cite{CN}, displays
\cite{CSA} and quantum dot lasers \cite{Klim}. Due to the unique polar axis of
hexagonal wurtzite geometry of these semiconductors, these structures often
form more anisotropic shapes \cite{MSA}. These higher aspect ratio polymorphs,
which lead to highly polarised emission, have many applications in
optoelectronics \cite{Empe}. Current synthesis techniques even allow for the
diversity of geometric structures such as nanowires, nanobelts, nanotubes and
nanorods. In particular, one dimensional structures like nanorods have shown to
support a high density of excitons and offer the possibility of enhanced
transport of dissociated charge carriers. Shape-controlled synthesis for CdSe
nanorods has been reported in last decade \cite{HYMAA, xpeng, Barnard}. \\ 

Metal-semiconductor (M-S) interfaces and nanostructured systems are also
gathering interest among the science community \cite{SMBM, dPdF, MSSRB, SSB,
CSESB} due to their potential applications in developing electronic and
optoelectronic devices. The space-charge region in bulk M-S interfaces extends
upto few nanometers.  It is therefore fascinating to study nanostructures which
physically are even less than few nanometers. The electronic properties of
nanoscale counterpart of bulk M-S interfaces are yet to be understood
completely. From the technological point of view, understanding the properties
of such nanocontacts is a major step in the route towards the implementation of
semiconductor nanocrystals (as well as molecules) in nanoelectronic device
architectures. With the aim of finding answers to the specific issues
pertaining to nanoelectronic device architectures, Landman et al studied the M-S
nanojunction problem theoretically \cite{Lan}. They showed induction of subgap
states near the Si-Al interface, decaying in Si nanowire and development of
relatively large Schottky barriers in comparison to bulk. Demechenko and Wang
calculated electronic structure of CdSe nanowires, in contact with metallic
electrodes of experimentally relevant sizes, by incorporating the electrostatic
image potential in atomistic single particle Schr\"{o}dinger equation \cite{DW}.
They demonstrated strong nanowire-size-dependence of localized electron-hole
states induced by the electrode. \\

In a scanning tunneling spectroscopy study of gold-tipped CdSe nanorods
(nanodumbbell), Steiner et al observed a gap similar to that in bare CdSe
nanorods near the nanodumbbell center, while subgap structure is found near the
M-S nanocontact \cite{SMBM}. They attributed this behaviour to the formation of
subgap interface states that vanished rapidly towards the center of the rod,
consistent with theoretical predictions given by Landman et al \cite{Lan}.
These states lead also to modified Coulomb staircase and in some cases to
negative differential conductance on the gold tips. The tunneling spectra can
be correlated with the theoretical calculation of density of states (DOS).
Theoretical predictions regarding range for decay of metal induced subgap
states is about 1~nm \cite{Lan} whereas experiments indicate that the subgap
states exist even up to 5~nm from the interface \cite{SMBM}. The experimental 
estimates are dictated by spatial resolution and the fact that locality of
tunneling spectra holds up to approximately the exciton-Bohr radius of material
\cite{SMBM} ($\sim$~5.6~nm for CdSe \cite{MLWZTB}). Present work is motivated by
this experimental work of Steiner et al.\\

In present work, we study passivated CdSe nanorods of various lengths in order
to see the effect of length variation on charge density profile of the rods. We
also study their density of states profile upon formation of nanocontacts with
gold clusters. We found that our results are qualitatively similar to those of
Landman et al \cite{Lan} and Steiner et al \cite{SMBM}. \\

\section{Computational details} 

Our calculations are based on density functional theory, implemented through
Vienna Ab-initio Simulations Package \cite{VASP} and are performed employing
plane augmented wave \cite{paw} with exchange-correlation energy functional as
given by Perdew, Burke and Ernzerhof \cite{pbe}. The valence electronic
configurations for Cd, Se and Au atoms are, $5s^{2}4d^{10}$, $4s^{2}4p^{4}$ and
$6s^{1}5d^{10}$ respectively. The cut-off energy used in plane wave expansion
is 274.34~eV. The self-consistent convergence of energy is set to 10$^{-5}$~eV.
The calculations are performed only at a single $k$-point, namely the center
of the Brillouin zone.  Occupation numbers are treated according to the
Fermi-Dirac scheme with a broadening of 0.001~eV. A sufficiently large unit
cell is chosen for the free standing CdSe nanorods so that the minimum distance
from the cluster boundary to unit cell boundary was 5~\AA\ in each direction.
This vacuum region is even larger (7~\AA) for gold attached CdSe nanorods. \\

To understand the structural stability, lattice-dynamical calculations
(resulting in phonon DOS) are performed within the framework of self-consistent
density functional perturbation theory. Force constant matrices thus obtained
are used to generate vibrational frequencies of the system. In order to analyse
the bonding between gold cluster and CdSe nanorods, Bader charge analysis
method is used based on the algorithm developed by Henkelman et al
\cite{bader1, bader2, bader3}. \\

\section{Results and Discussions}
\subsection{Cluster derived CdSe nanorods}
\begin{figure}[ht]
\begin{center}
	\includegraphics[width=8.5cm]{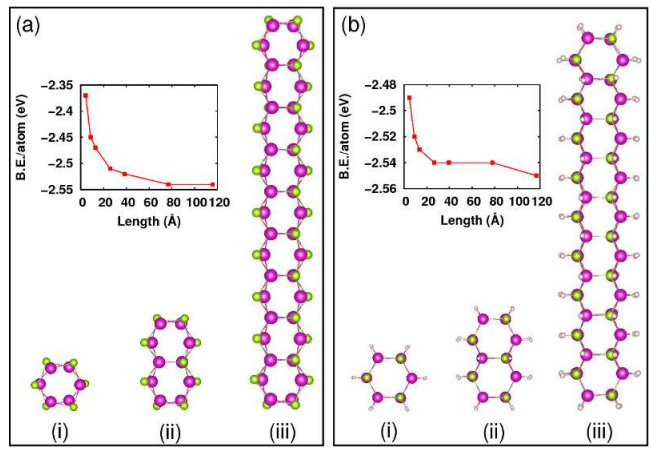} \\
\end{center}
\caption{(color online) Geometries of (a)~bare and (b)~passivated nanorods of various sizes. 
(i)~Cd$_6$Se$_6$ (4.3~\AA, 4.6~\AA), (ii)~Cd$_{10}$Se$_{10}$ (8.7~\AA, 9.0~\AA) and 
(iii)~Cd$_{38}$Se$_{38}$ (38.6~\AA, 39.4~\AA). Numbers in brackets indicate 
the lengths of the corresponding bare and passivated nanorods respectively. 
Cd, Se and H* atoms are indicated in magenta, green and pink respectively. The 
colour scheme is maintained throughout this article. Insets show the variation of BE 
per atom against lengths of the nanorods. Note the difference in the BE per atom scale 
on $y$-axis in (a) and (b).
\label{fig:1}} 
\end{figure} 

The nanorods are generated by relaxing a fragment (viz. Cd$_3$Se$_3$) of bulk
wurtzite structure of CdSe crystal in $[\bar{1}~0~1~0]$ direction, into its
global minimum. This fragment has planar structure and can be considered as the
basic building block of the longer nanorods (Cd$_n$Se$_n$). In this cluster, Cd
atoms form an equilateral triangle of side 3.22~\AA\ and Cd-Se bond length is
2.51~\AA. Further, Cd$_6$Se$_6$ nanorod is constructed by stacking these six
atom rings in chair conformation (See Figure \ref{fig:1}-a(ii)). This is the
smallest nanorod under consideration. The atoms in this geometry relax in such
a manner that intraplanar Cd-Se bond lengths are optimized to 2.60~\AA, and
Cd-Se interplanar bond lengths become 2.82~\AA. The length of this nanorod is
found to be 4.34~\AA. The longer nanorods are then constructed as the growth
over Cd$_6$Se$_6$ cluster and can be viewed as ``n'' Cd$_6$Se$_6$ clusters
coupled together sharing a Cd$_2$Se$_2$ in common (See Figure \ref{fig:1}).
Thus, we are dealing with the structures where all the atoms reside on the
surface. Binding energy (BE) (defined as the difference between total energies
of nanorods and the constituent atoms) indicate that structure is stable. As
shown by the graph in the inset of Fig. \ref{fig:1}(a), BE per atom reduces
initially with the length of the nanorod and then saturates (See also Table
\ref{table:1}). At smaller lengths of the nanorods, the structures are more or
less spherical. But as the length increases the aspect ratio increases, thus
making the BE per atom constant.\\

\begin{table}
\caption{Comparison between the results for passivated and unpassivated nanorods.
\label{table:1}}
\begin{tabular}{|c| c c | c c | c c| }
	\hline
        {~}	& \multicolumn {2}{|c|}{~}	& \multicolumn {2}{|c|}{~} & \multicolumn {2}{|c|}{~} \\
        {\bf Structures}	& \multicolumn {2}{|c|}{{\bf Length (\AA)}}	& \multicolumn {2}{|c|}{{\bf Band gap}} & \multicolumn {2}{|c|}{ {\bf BE per}} \\
                        & \multicolumn {2}{|c|}{}                               & \multicolumn {2}{|c|}{{\bf (eV)}}              & \multicolumn {2}{|c|}{ {\bf atom (eV)}} \\
	\cline {2-7}
	& ~Bare~  &  +H* &  ~Bare~ & +H* &  ~Bare~ & +H* \\
        \hline \hline
        Cd$_6$Se$_6$    & 4.34	& 4.64	& 2.02  & 3.17 & -2.37 & -2.49  \\
			&	&	&	&	&	& \\
	Cd$_{10}$Se$_{10}$    & 8.65	& 9.04	& 1.98 & 2.99  & -2.45  & -2.52 \\
			&	&	&	&	&	& \\
	Cd$_{14}$Se$_{14}$    & 12.91	& 13.50	& 2.03 & 2.94  & -2.47  & -2.53   \\
			&	&	&	&	&	& \\
	Cd$_{26}$Se$_{26}$    & 25.76	& 26.60	&  1.99 & 2.71  & -2.51  & -2.54     \\
			&	&	&	&	&	& \\
	Cd$_{38}$Se$_{38}$    & 38.55	& 39.43	&  1.95 & 2.52  & -2.52 & -2.54   \\
			&	&	&	&	&	& \\
	Cd$_{74}$Se$_{74}$    & 76.96	& 77.78	&  1.88 & 2.42  & -2.54 & -2.54     \\
			&	&	&	&	&	& \\
	Cd$_{110}$Se$_{110}$    & 115.85	&  116.66	& 1.87  & 2.35  & -2.54  & -2.55     \\
        \hline

\end{tabular}
\end{table}

Surface atoms, due to lower co-ordination number, have incomplete bonding
resulting in electronically active states, known as ``dangling or unpassivated
orbitals''. These orbitals are localized and act as efficient traps for charge
carriers. At nano-scale, materials become very sensitive to the surface
properties due to high surface to volume ratio. Hence, surface passivation is
essential for such semiconductor nanostructures. This removes the localized
surface states from the band gap.  It is experimentally evident that for the
II-VI group heteropolar semiconductor nanostructures like CdSe;
trioctylphosphine oxide (TOPO) or trioctylphosphine (TOP) are the right choice
as passivating agents due to their optimal bonding to the nanostructure surface
\cite{Ali, Hea, YTMI}. The complex and large atomic structures of these
passivating agents are costly for computational calulations. Hence, one needs
to explore simpler computational methods for surface passivation. There are
various techniques developed for such calculations and each one has its own
advantages and disadvantages \cite{WZ, Shi, WL}. For our calculations we use
the technique developed by Huang et al \cite{HLC}, where fictitious hydrogen
atoms, H*, are chosen to complete the co-ordination of the surface atoms. The
pseudopotentials for these pseudo-atoms with appropriate nuclear and valence
charge, are generated using the Troullier-Martins prescription \cite{TM}. The
authors suggest that CdSe dots with a wurtzite structure open a maximum band
gap when H* atoms with valence electron charge Z = 1.5 are used to bond with Cd
atoms (cations) and Z = 0.5 to bond with Se atoms (anions), thus maintaining
the charge neutrality of the whole system. For passivated CdSe nanorods, the
wurtzite symmetry is maintained (See Figure \ref{fig:1}(b)), but Cd-Se bond
length is now elongated upto 2.71~\AA\ and lengths increase slightly (maximum
upto $\sim$~0.9~\AA) with respect to the non-passivated nanorods. Cd atoms bond
with H* atoms with bond length 1.84~\AA\ while Se-H* bond length is 1.60~\AA.
The structural stability of these nanorods is also confirmed using vibrational
frequency analysis. Absence of imaginary frequencies establishes that the
geometries of these nanorods are locally stable. \\

\begin{figure}
\begin{center}
\includegraphics[width=6.25cm]{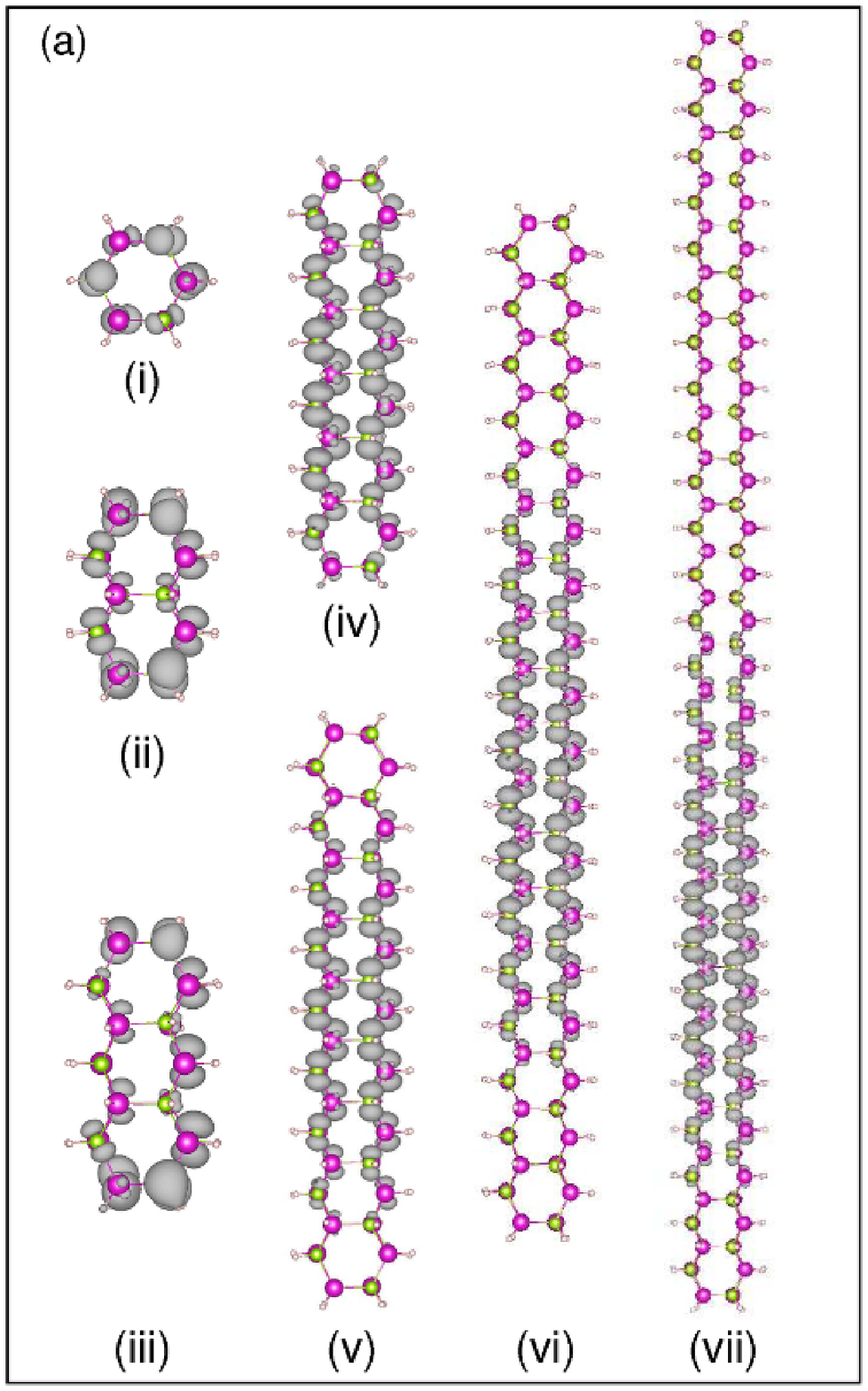} \\~\\
\includegraphics[width=6.25cm]{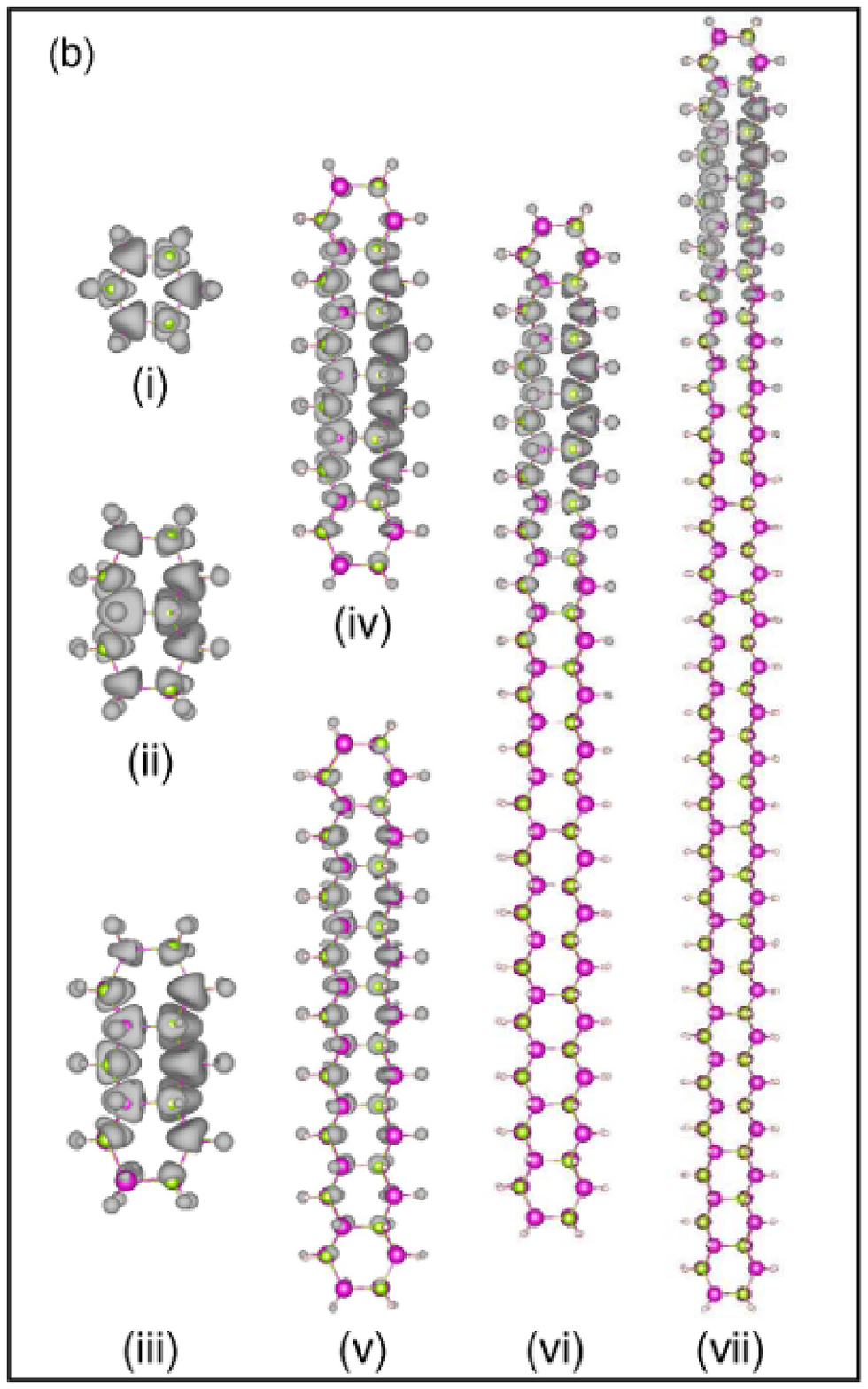} \\~\\
\end{center}
\caption{Partial charge densities of the nanorods showing states HOMO in (a)
and LUMO in (b). Note the presence of $p$-states in HOMO on Se atoms while $s$
as well as $p$-states in LUMO on Cd and Se atoms. As the length of the rods
increases, there is a growing asymmetry in the distribution of these states
whose span saturates beyond a certain length.
\label{fig:2}}
\end{figure}

Table \ref{table:1} gives the results of the calculations for passivated and
unpassivated CdSe nanorods of different lengths. BE per atom for passivated
structures indicates the same trend as that of non-passivated structures (See
inset of Fig. \ref{fig:1}(b)), but is lower for passivated structures with low
aspect ratio. Thus, passivation takes care of the surface effects prevalent in
structures of lower aspect ratios. A comparison of local density of states
(LDOS) of these structures also indicates that passivation does not change the
intrinsic behavior of highest occupied molecular orbitals (HOMO) and lowest
unoccupied molecular orbitals (LUMO). Hence, we could safely conclude that
surface passivation is essential in case of such CdSe one dimensional
structures. Inspection of BG values of passivated nanorods shows a significant
increase by upto 1.1~eV for the smallest length of nanorod. Thus, surface
passivation removes the localized surface states from the gap region and these
results contradict ``self-healing'' reported by Puzder et al \cite{PWGG} in
their study of CdSe clusters in the form of wurtzite cages of different
diameters. In their work , Puzder et al observe that the surface relaxation in
CdSe nanostructures act in similar manner to that of passivated nanostructures
of CdSe by opening the gap substantially. \\

\begin{figure}
\begin{center}
\includegraphics[width=9cm]{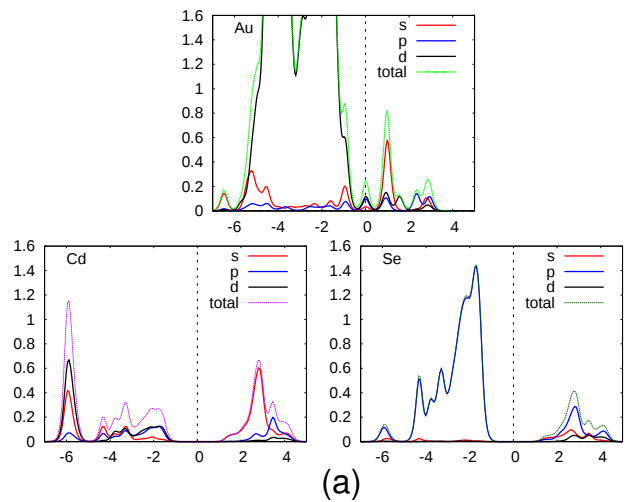}
\end{center}
\caption{Site projected DOS of Au atom in gold cluster (Au$_{13}$) and Cd and Se
atoms in free standing passivated nanorods (Cd$_n$Se$_n$). Note the dominant presence of $p$ states of Se in HOMO and $s$ and
$p$ states of Cd as well as Se in LUMO.
\label{fig:3}}
\end{figure}

Figures \ref{fig:2}(a) and (b) show the partial charge density plots for HOMO
and LUMO of passivated CdSe nanorods respectively. The partial charge density
distribution of HOMO is predominantly localized on Se atoms which comes from
the $p_y$ orbitals (which lie along the length of the rod) of the atoms while
that of LUMO is on both Cd and Se atoms arising from $s$ orbitals of both Cd
and Se throughout the nanorod. This is also reflected in the site projected DOS
(See Figure \ref{fig:3}). For short nanorods, (\textless 15~\AA), $p_z$
orbitals of Se atoms at the edges also contribute to HOMO. The range of partial
charge density arising from HOMO saturates at 39~\AA\ for longer rods.  In case
of LUMO, as the length of the nanorods increases, the contribution to the
partial charge density from Cd atoms, goes on decreasing. The span of LUMO
saturates over the length of 30~\AA.  The passivating agents show the
contribution from $s$ orbitals in the LUMO of the nanorods. Bader charge
analysis shows that the bonding between Cd and Se atoms in the nanorods has a
partial ionic character with Cd transferring an average charge of 0.55$e$ on
Se.\\

\subsection{Gold tipped CdSe nanorods}

\begin{figure} 
\begin{center} 
\includegraphics[width=4cm]{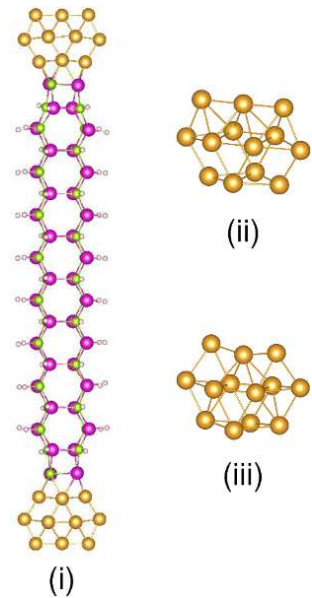}
\end{center} 
\caption{(i) Gold (Au$_{13}$) tipped passivated Cd$_{38}$Se$_{38}$
nanorod. Au$_{13}$ clusters : (ii) bare ; (iii) upon attaching at the tips of
CdSe nanorods. Notice the buckling in the geometry upon attaching at the tips
of the rods. Gold atoms at the interface are bonded to Cd as well Se atoms at
the tips of the rods.  
\label{fig:4}} 
\end{figure}

Global minimum geometry of Au$_{13}$ cluster is found to be suitable to attach
at the ends of the nanorods. In this cluster, gold atoms are arranged in three
distinct planes. The bare clusters are attached to the fully relaxed nanorods.
These compound structures are then optimized fully forming a dumbbell-like
structure (nanodumbbell). We observe that the nanorods are unaffected due the
end passivation by gold electrodes, except at the M-S nanojunctions, where CdSe
bond length elongates to increase the area of the quadrilateral formed by two
CdSe pairs. Geometry of gold cluster however changes significantly (See Figures
\ref{fig:4}(ii) and (iii)). We can observe a slight buckling in three planes of
the cluster towards M-S interface.  Both Cd and Se atoms at the ends of the
nanorods are doubly bonded to the Au atoms at the junctions, where Au-Cd and
Au-Se average bond lengths are 2.74~\AA\ and 2.85~\AA\ respectively. \\

\begin{figure} 
\begin{center} 
\includegraphics[width=8.5cm]{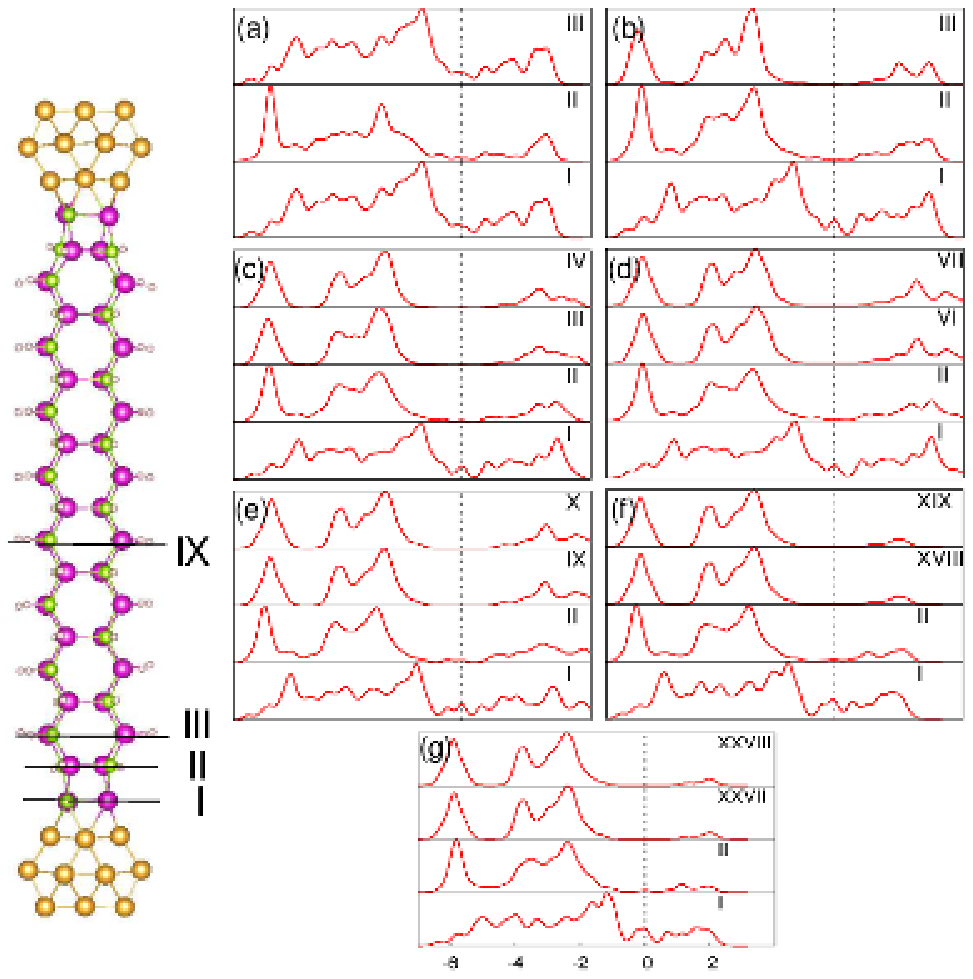}
\end{center}
\caption{(color online) LDOS corresponding to various planes, as
depicted on the left for gold (Au$_{13}$) tipped passivated Cd$_{38}$Se$_{38}$
nanorod, are shown for following nanorods (Au atoms are indicated in yellow):
(a)~Cd$_6$Se$_6$, (b)~Cd$_{10}$Se$_{10}$, (c)~Cd$_{14}$Se$_{14}$,
(d)~Cd$_{26}$Se$_{26}$, (e)~Cd$_{38}$Se$_{38}$, (f)~Cd$_{74}$Se$_{74}$ and
(g)~Cd$_{110}$Se$_{110}$. The Fermi level (shifted to zero of energy) is
marked by the vertical dotted line. Each plane in the nanorod contains two CdSe
pairs along with their passivating H* atoms. Numbering is done only up to the
central plane owing to the symmetry of the geometry as well as the electronic
structure of nanorods (as is evident in (a) for planes I and III). LDOS shows
MIGS present at the central plane in structures (a), (b),
(c) and (d). These states are absent around the central planes of (e), (f) and
(g). The gap between HOMO and LUMO increases as we approach the central plane and saturates for longer rods. 
\label{fig:5}} 
\end{figure}

\begin{figure} 
\begin{center}
\includegraphics[width=7.5cm]{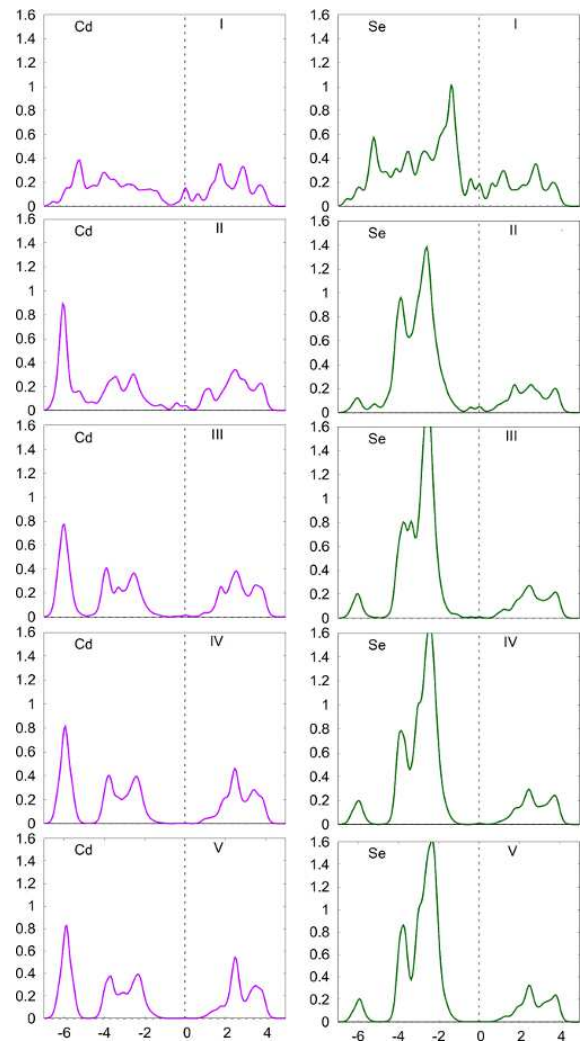} \\   
\end{center}
\caption{Plane-wise site projected DOS of Cd atoms in left column and Se atoms in right, in gold tipped
Cd$_{38}$Se$_{38}$ rod. In the plane near the junction, one can see the
presence of gold states in the gap. These states are also present over a wide
energy scale below HOMO, indicating hybridization. Gold states are drastically
reduced in the very next plane. As we go towards the center of the rod, the DOS
does not show the presence of gold states and the DOS profile of Cd and Se
looks the same as that of free standing nanorod.  
\label{fig:6}} 
\end{figure}

LDOS calculated for every plane of the gold tipped nanorods (See Figure
\ref{fig:5}), shows a gap near nanodumbbell center and metal induced gap states
(MIGS) emerge near M-S nanocontact. These MIGS vanish rapidly towards the
center of the nanorod \cite{SMBM, Lan}. For short nanorods (upto
$\sim$~27~\AA) bridging gold electrodes, we find full metallization by these
MIGS (Fig. \ref{fig:5}(a), (b), (c) and (d)), while for longer nanorods, we
find a gap-structure emerging away from M-S junction. For longer nanorods, in
interfacial regions closest to the metal electrodes, a finite LDOS is observed
in the gap indicating a metallic nature which gradually decays away from the
electrode. A comparison of site projected DOS for gold attached nanorods (See
Figure \ref{fig:3}) and their separated components, shows the hybridization of
metal states with semiconductor states. There is maximum opening of the gap at
the central plane of nanodumbbell and the gap size decreases with increasing
length of the nanorod, eventually saturating to a value 2.05~eV. Figure
\ref{fig:6} shows that the plane-wise site projected LDOS for
Au$_{13}$Cd$_{38}$Se$_{38}$Au$_{13}$ nanodumbbell where MIGS start vanishing rapidly across the
planes and central plane of the rod shows the presence of only semiconductor
states. MIGS are present only upto a distance of $\sim$~15.5~\AA\ from the
interface, thus making it clear why the nanorods shorter than 27~\AA\ show full
metallization.\\

\begin{figure} 
\begin{center} 
\includegraphics[width=9cm]{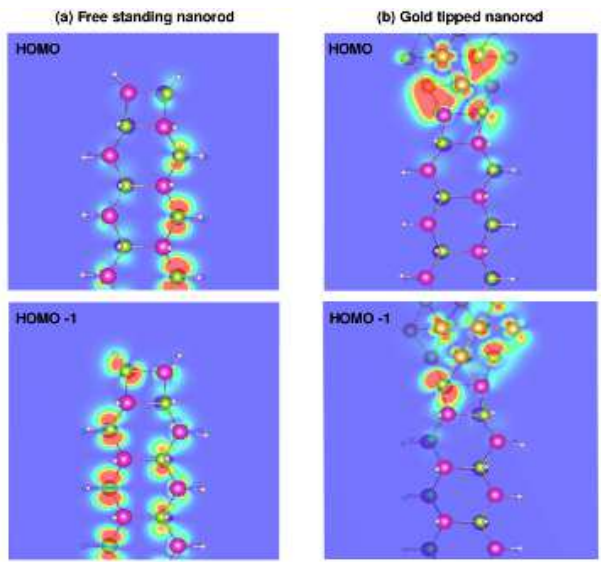}
\end{center} 
\caption{Partial charge density contour plots for HOMO and HOMO-1 states of 
(a) free standing and (b) gold tipped passivated Cd$_{38}$Se$_{38}$ nanorod. 
\label{fig:7}} 
\end{figure} 

Figure \ref{fig:7} depicts the partial charge density plots for HOMO
and HOMO-1 states for free standing and gold attached passivated
Cd$_{38}$Se$_{38}$ nanorod. For the nanodumbbells, the HOMO and HOMO-1 states
are comprised of gold $d$ and $s$ states hybridized with $p_z$ states of Se and
$s$ states of Cd as shown in Fig. \ref{fig:7}(b). This behaviour is also
evident from the LDOS plots in Fig. \ref{fig:6}. Hybridization of gold
states with Cd states is not evident in the partial charge density plots.
However, gold induced states appear around Cd atoms in the contact plane (See
Fig. \ref{fig:6} - Plane I and II). The $p_y$ states of Se atoms comprising
the HOMO and nearby levels in free standing passivated Cd$_{38}$Se$_{38}$
nanorod have shifted below Fermi level ($E_F$) by $\sim$2~eV, while states near
HOMO in gold tipped nanorod mainly arise from gold and are spatially localized
near the contact plane. Bader charge analysis for the nanodumbbells shows that
changes in the charge transfer in comparison to free standing rods are
localized near the interface. Bonding remains same away from the interface. At
the interface, Cd atoms transfer an average charge of 0.68$e$ while Se atoms
acquire an average charge of 0.53$e$. The remaining charge gets distributed on
the gold cluster. \\

\begin{table}
\caption{Schottky barrier height (SBH) and HOMO-LUMO gap at the central plane as a function of the length
of CdSe nanorods. SBH for bulk gold-CdSe contacts is 0.7~eV at 300~K \cite{sze}.  
\label{table:2}} 
\begin{tabular}{| c | c | c | c | } 
        \hline
        {~}                     &       &                               &  \\
        {\bf Structures}        & {\bf Length}  & {\bf Schottky barrier}     & {\bf Gap at} \\
                                &               & {\bf height}      & {\bf central plane} \\
                                & {\bf (\AA)}   & {\bf (eV)}    & {\bf (eV)}            \\
        \hline \hline 
        Cd$_{38}$Se$_{38}$      & 39.43 & 1.00          & 2.12               \\
        Cd$_{74}$Se$_{74}$      & 77.78 & 0.92          & 2.06              \\
        Cd$_{110}$Se$_{110}$    & 116.66 & 0.93         & 2.05             \\
        \hline
\end{tabular}
\end{table}

Difference in work function of metal (Au) and electron affinity of
semiconductor (CdSe) yields a Schottky barrier when the two are brought in
contact. A decrease in the number of available states in confined structures
causes an increase in the energy barrier that carriers have to surmount in
order to cross the interface. Quantum confinement of carriers is known to
increase the minimum energy that a carrier has to have (relative to HOMO for
electrons and LUMO for holes) to propagate. \\ 

For an n-type semiconductor like CdSe, the Schottky barrier height (SBH) is
given by the difference between the metal work function and the electron
affinity of the semiconductor. This is calculated from the position of $E_F$ in
the nanodumbbell and the position of LUMO in the middle section of the CdSe
nanorod following Landman et al \cite{Lan}. The values for SBH are listed in
Table \ref{table:2}. Group II-VI semiconductors have a higher component of
ionic bonding and hence these materials do not create large number of surface
states. Their barrier heights therefore depend upon the work function of the
metal. Here, the SBH for longer nanorods, which do not show metallization,
saturate for the nanorods having lengths longer than $\sim$78~\AA. \\

\section{Conclusions} 

In summary, we report that passivation of CdSe nanorods opens up the band gap
considerably and we do not observe any ``self-healing'' as reported by Puzder
et al \cite{PWGG} for their 3D structures. For free standing nanorods contribution
to HOMO comes mainly from $p$ orbitals of Se similar to bulk and 3D confined
structures. However contribution to LUMO comes from $s$ orbitals of Cd and Se
while for bulk and 3D structures LUMO mainly consists of Cd $s$ states. HOMO is
confined over a region of 39~\AA, while LUMO is confined over 30~\AA. Gold
attached nanorods are fully metalized for shorter lengths (\textless~27~\AA),
while they develop a Schottky barrier, larger than the bulk value (0.7~eV at
300~K\cite{sze}), for longer nanorods, where a semiconducting band gap starts to show
up at a distance of 15.5~\AA\ from the nanojunction. End-passivated nanorods
and their separated components show charge transfer which is highly localized
at contact region.\\ 

All the figures of the structures are generated using VESTA \cite{vesta}. 
We thank Department of Science \& Technology, Government of India for financial 
support and C-DAC, Pune for use of their computing facilities.

\bibliography{biblio}
\end{document}